\documentclass[twoside]{ae100prg}
\usepackage{graphicx,subfig,mathrsfs,amsmath}
\usepackage[breaklinks]{hyperref}
\usepackage{booktabs}

\newcommand{\IG}[2]{{\includegraphics[keepaspectratio,width=#1\textwidth]{#2.eps}}}
\newcommand{\re}[1]{(\ref{#1})}
\renewcommand{\d}{\mathrm{d}}
\graphicspath{{./pictures/}}

\begin{document}
\title{Variations on spacetimes with boost-rotation symmetry}
\author{Ji\v{r}\'{i} Bi\v{c}\'ak$^1$, David Kofro\v{n}$^1$}

\address{$^1$ Institute of Theoretical Physics,\\
Charles University in Prague, Czech Republic}

\email{bicak.troja@gmail.com, d.kofron@gmail.com}

\begin{abstract}
Some new results on the boost-rotation symmetric spacetimes representing pairs of rotating charged objects accelerated in opposite directions are summarized. A particular attention is paid to (a) the Newtonian limit analyzed using the Ehlers frame theory and (b) the special-relativistic limit of the C-metric.  
Starting from the new, simpler form of the rotating charged \mbox{C-metric} we also show how to remove nodal singularities and obtain a rotating charged black hole freely falling in an external electromagnetic field.

\end{abstract}

\section{Boost-rotation symmetric spacetimes}
Boost-rotation/axial symmetric spacetimes are important explicit examples of exact solutions of Einstein field equations describing non-trivially moving sources of gravitational and electromagnetic field \cite{bischm}. The only ``initial'' assumption that is made is the existence of two Killing vectors: boost Killing vector ${\xi^{\mu}_B}$ whose orbits are hyperbolas and axial Killing vector ${\xi^{\mu}_{\phi}}$ with closed circular orbits. The metric of a general electrovacuum rotating boost-rotation symmetric spacetime in global coordinates reads:
\begin{equation}
\d s^2 = \frac{e^{\mu}\left(z\d t-t\d z+ \Omega\d\phi\right)^2-e^{\nu} \left(z\d z-t\d t\right)^2}{z^2-t^2} - e^{\nu}\d r^2-e^{-\mu}r^2\d \phi^2\,. \label{eq:br}
\end{equation}
Functions $\mu$, $\nu$ and $\Omega$ depend on $a=z^2-t^2$ and $b=r^2$ only and are determined by the Ernst equations (basic consequence of the Einstein field equations under these symmetries) and the character of sources. The nonlinear Ernst equations are difficult to solve but if rotation $\Omega$ and electromagnetic field vanish, then the basic Einstein field equation reduces to the wave equation on an auxiliary flat spacetime $\square \mu=0$ while $\nu$ can be determined by quadrature.

\begin{figure}
\begin{center}
\hspace*{\fill}\begin{minipage}{.4\textwidth}
This diagram schematically indicates the world lines of sources (thick lines) and different conicity regions of the axis (grey), or world-sheets of the black-holes horizons.
\end{minipage}\hfill
\begin{minipage}{.4\textwidth}
\IG{1}{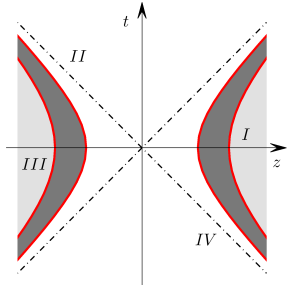}
\end{minipage}\hspace*{\fill}
\end{center}
\caption{Space-time diagram of boost-rotation symmetric solution.}
\vspace{-1em}
\label{fig:STdiag}
\end{figure}

The main features of the boost-rotation symmetric spacetimes are:\\ 
\parbox{.05\textwidth}{(a)} They describe uniformly accelerated sources\\
\parbox{.05\textwidth}{(b)} \parbox[t]{.95\textwidth}{are asymptotically flat at null infinity except, in general, at two its generators}\\
\parbox{.05\textwidth}{(c)} \parbox[t]{.95\textwidth}{the hypersurfaces $z^2=t^2$, where the boost Killing vector ${\xi^\mu_B}$ is null, invariantly divide the spacetime into four quadrants (see Fig. \ref{fig:STdiag}) -- \\
\parbox{.07\textwidth}{(c1)} below the roof (reg. I and III) -- locally Weyl metrics\\
\parbox{.07\textwidth}{(c2)}
\parbox[t]{.87\textwidth}{above the roof (reg. II and IV) -- locally Einstein-Rosen, resp. Gowdy metrics}}\\
\parbox{.05\textwidth}{(d)} \parbox[t]{.95\textwidth}{are of algebraic type $I$, in general; the C-metric, describing accelerated black holes, is of type $D$}\\
\parbox{.05\textwidth}{(e)} are radiative with a non-vanishing news function\\
\parbox{.05\textwidth}{(f)} \parbox[t]{.95\textwidth}{along the axis of symmetry there are conical singularities in general -- they can be interpreted as strings or struts that cause the acceleration.}

\section{Newtonian limit}
Newtonian limit of a relativistic spacetime greatly corroborates its physical interpretation. 
We perform the limit within the framework of the Ehlers frame theory (see \cite{BiKofNL} for more details and references). The key point is the causality constant $\lambda=c^{-2}$ which becomes zero in the limit. To do the limit we have to choose a suitable set of observers and a naturally adapted coordinate system.

\begin{figure}[h]
\begin{center}
\hspace*{\fill}\subfloat{\IG{.28}{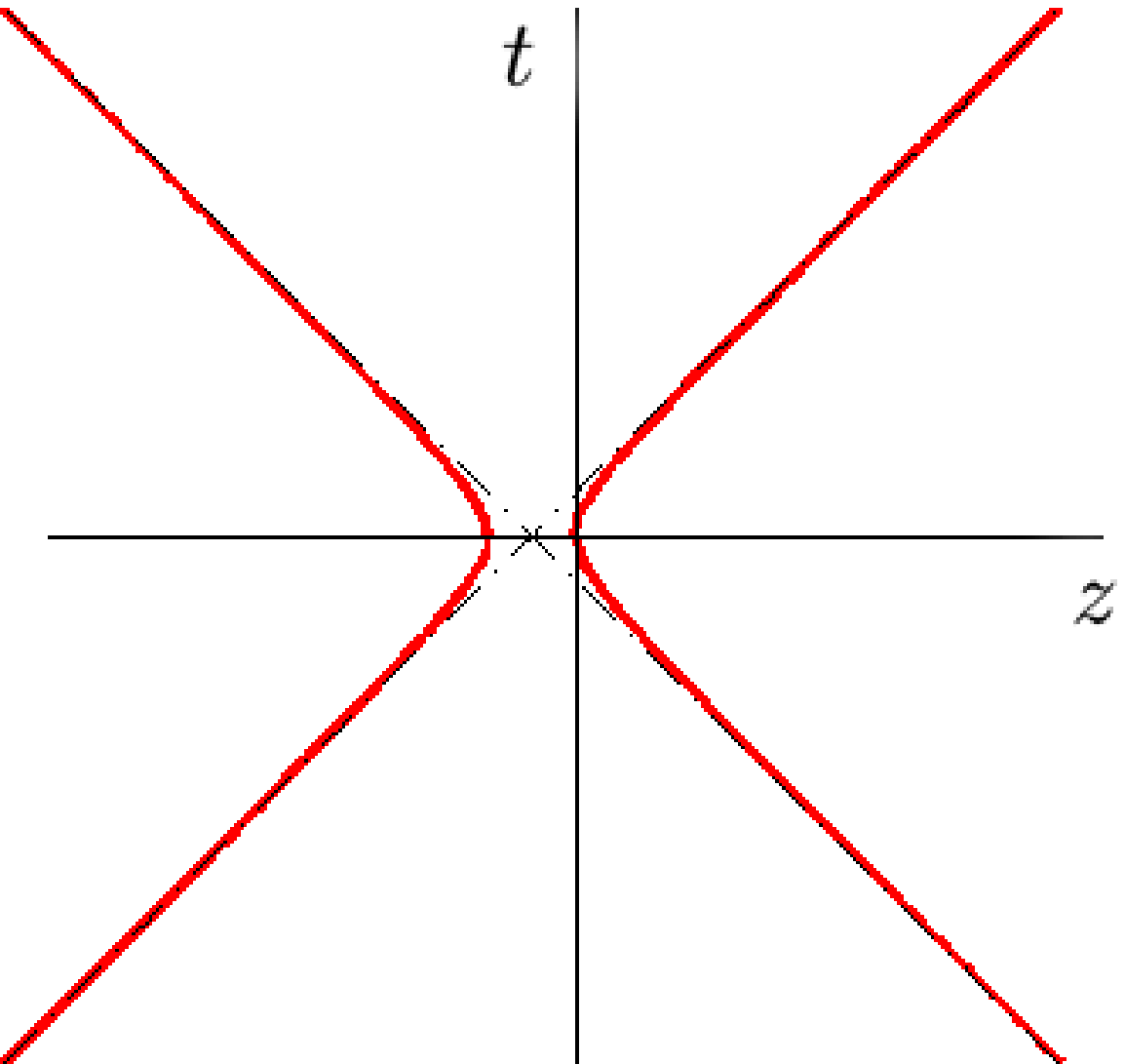}}\hfill
\subfloat{\IG{.28}{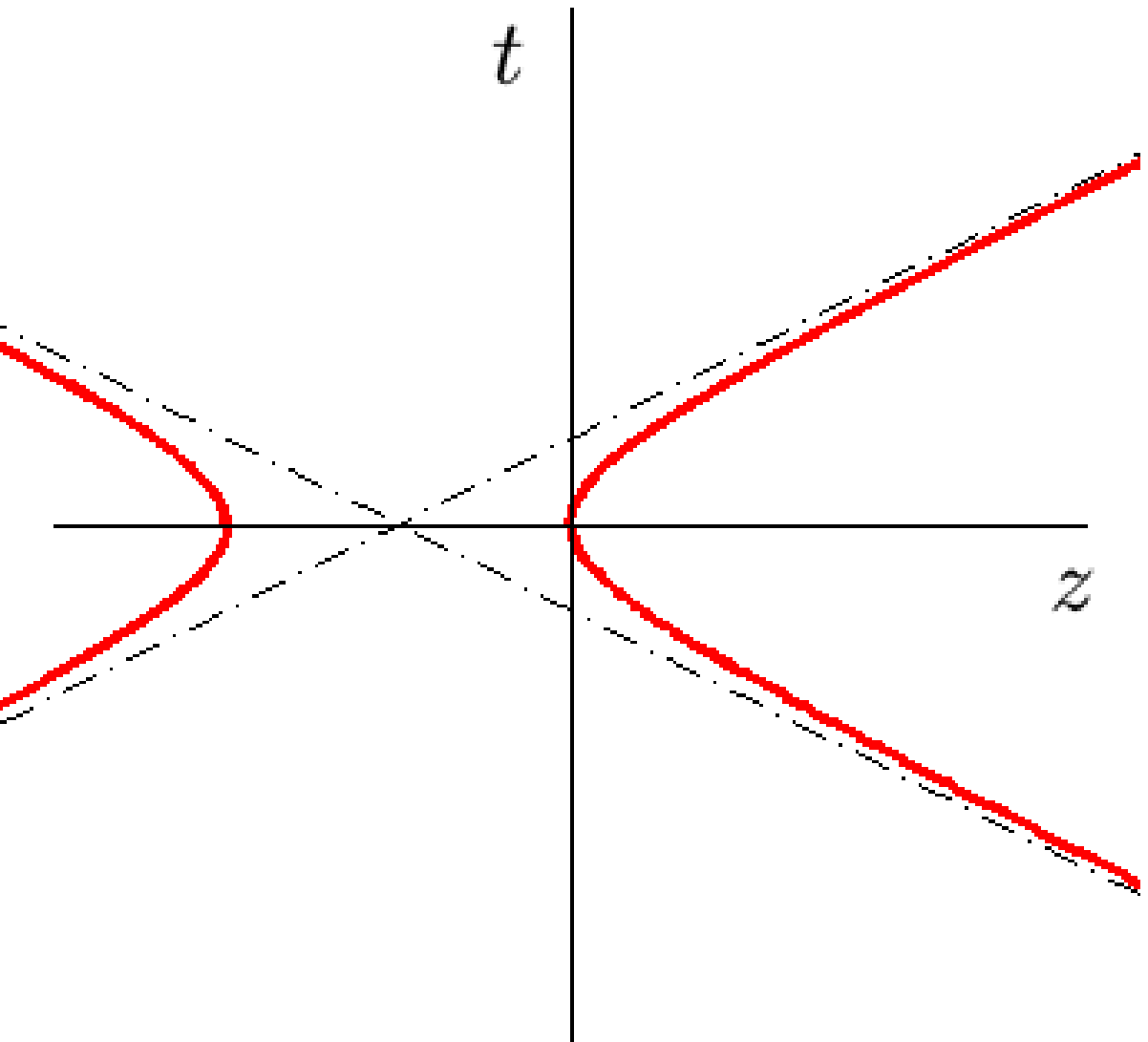}}\hfill
\subfloat{\IG{.28}{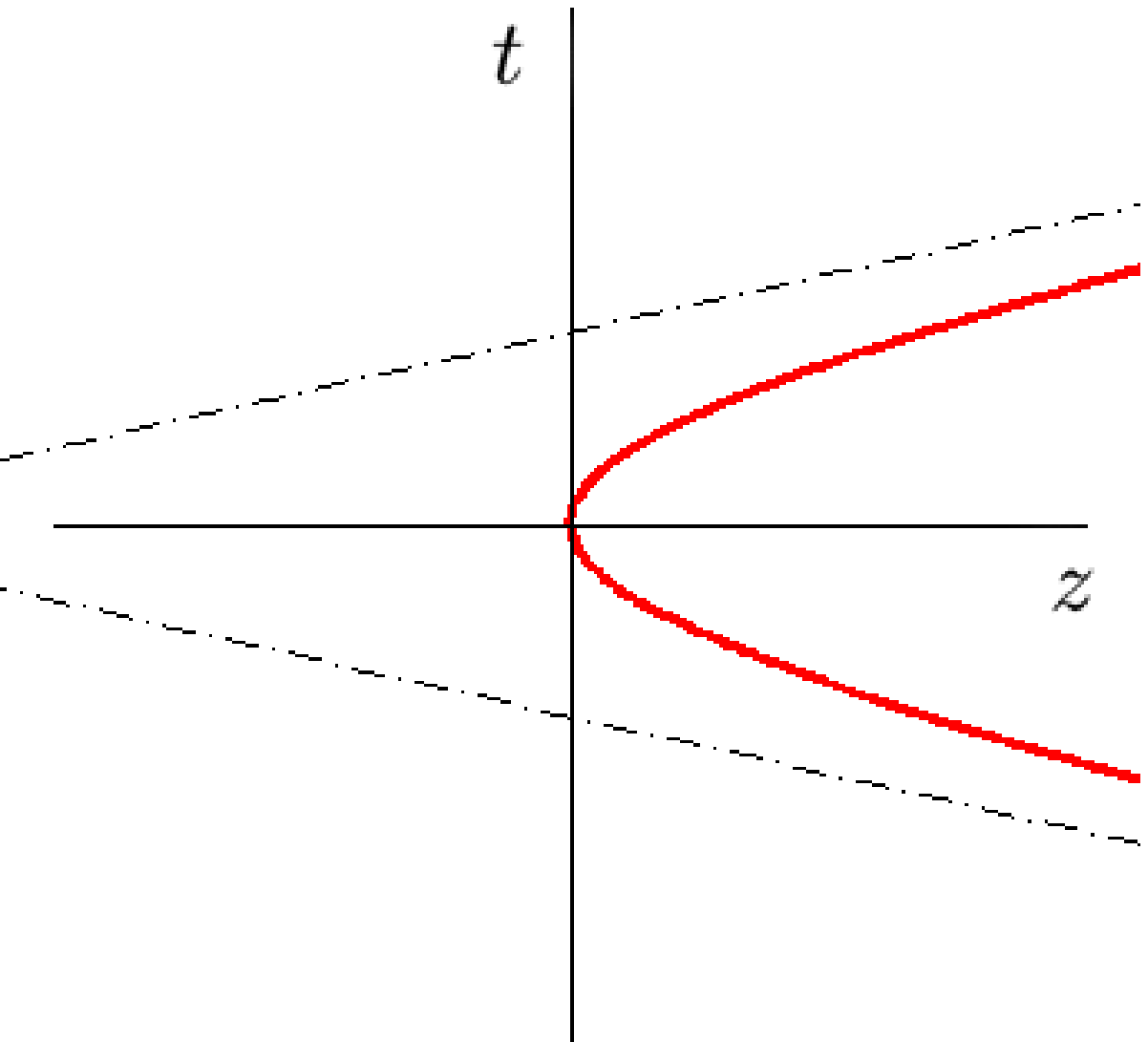}}\hspace*{\fill}\\
\hspace*{\fill}\subfloat[$\lambda=1$]{\IG{.28}{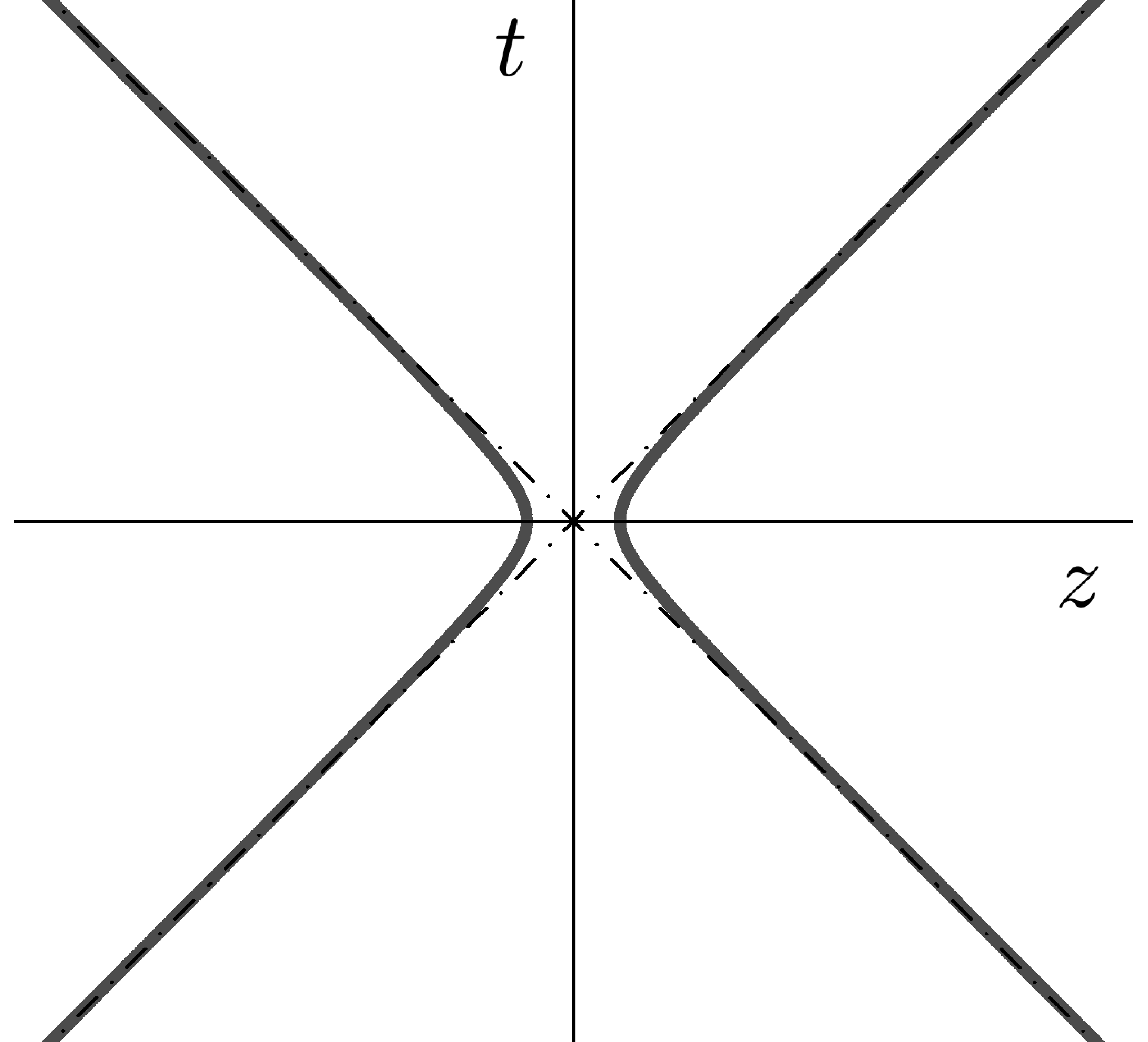}}\hfill
\subfloat[$\lambda=0.25$]{\IG{.28}{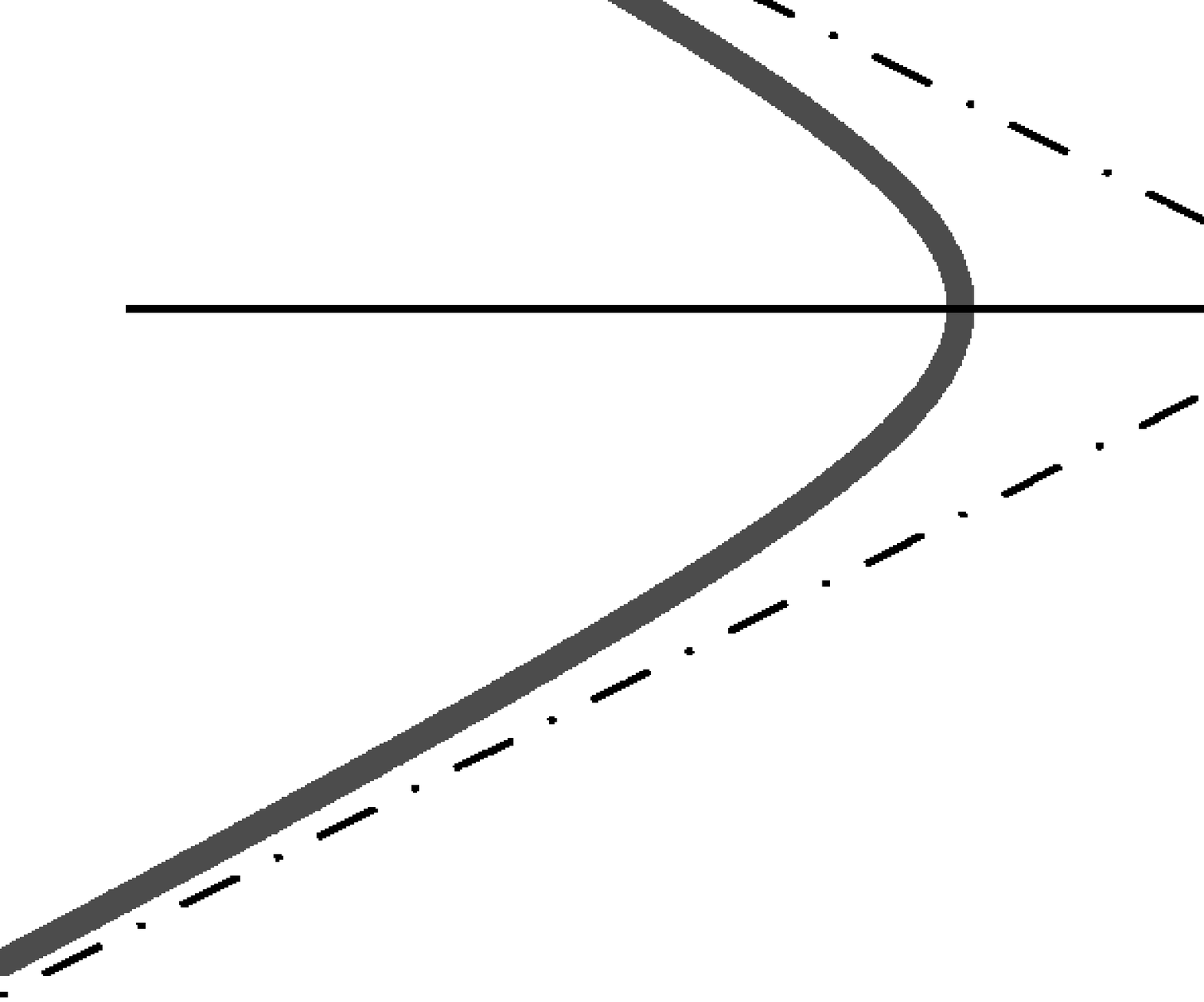}}\hfill
\subfloat[$\lambda=0.05$]{\IG{.28}{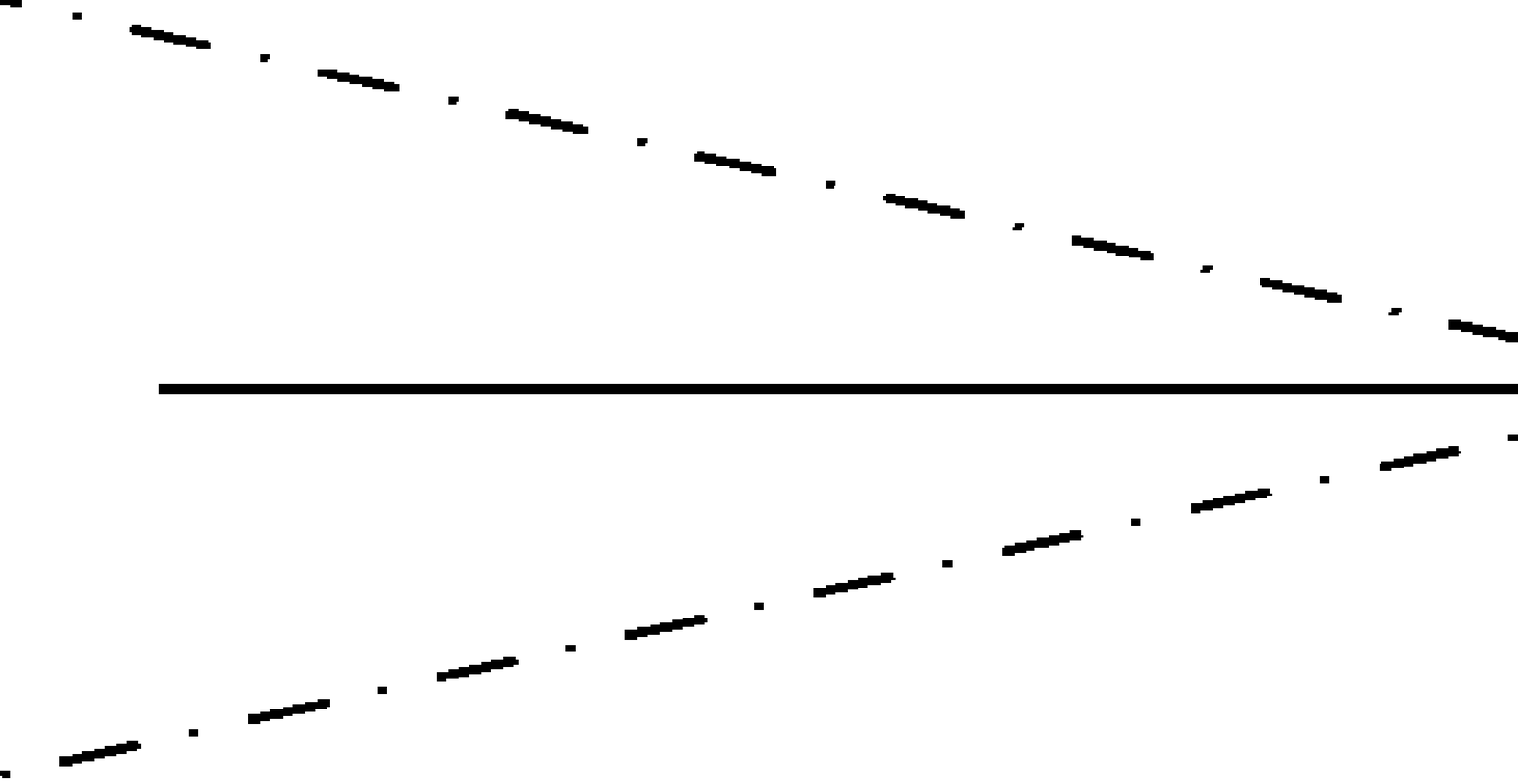}}\hspace*{\fill}\\
\end{center}
\caption{The sequence of spacetime diagrams with decreasing $\lambda=c^{-2}$ when making the coordinate shift (upper part) and without it (lower part). The worldlines of the sources are shown.}
\label{fig:NL}
\end{figure}

Using \re{eq:br} with functions $\mu$ and $\nu$ known, we first introduce a shift of the coordinate origin by putting $z=\zeta+\lambda^{-1} g^{-1}$; otherwise the particles would ``disappear'' to infinity (see Fig. \ref{fig:NL}).
Our procedure then results into a classical point particle undergoing uniform acceleration, $z=\frac{1}{2}gt^2$, which generates classical field described by the Newtonian gravitational potential $\Phi=\frac{m}{\sqrt{r^2+(z-\frac{1}{2}gt^2)^2}}$. This follows from the limit of various examples of these spacetimes \cite{BiKofNL}.

Our results thus strongly support the physical significance of the boost-rotation symmetric spacetimes (in contrast to some previous conclusions \cite{lazkoz-2004-460} which made the limit in the regions $II$ and $IV$).

\section{The rotating charged C-metric}
The rotating charged C-metric is a special case of the boost-rotation symmetric spacetimes. It describes two rotating charged black holes.
\begin{footnotesize}
\begin{multline}
\d s^2 = \frac{1}{A^2(x-y)^2}\;\biggl\{\; \frac{\mathcal{G}(y)}{1+\left( aAxy \right)^2}\;\Bigl[ \d t+aA\left( 1-x^2 \right)K\d\phi \Bigr]^2-\frac{1+\left( aAxy \right)^2}{\mathcal{G}(y)}\,\d y^2 \\
+\frac{1+\left( aAxy \right)^2}{\mathcal{G}(x)}\,\d x^2 +\frac{\mathcal{G}(x)}{1+\left( aAxy \right)^2}\;\Bigl[ \left( 1+a^2A^2y^2 \right)K\d\phi+aAy^2\d t \;\Bigr]^2\;\biggr\}\,.
\label{eq:rCM}
\end{multline}
\end{footnotesize}

\noindent Function $\mathcal{G}(\xi)$ is a polynomial which has recently been factorized by Emparan, Hong and Teo. The transformation between the forms \re{eq:rCM} and \re{eq:br} is explicitly known.

\begin{figure}
\begin{center}
\hspace*{\fill}\begin{minipage}{.4\textwidth}
The spacetime diagram of an accelerated magic electromagnetic field with the worldsheet of its source (the charged rotating disc).
\end{minipage}\hfill
\begin{minipage}{.4\textwidth}
\IG{1}{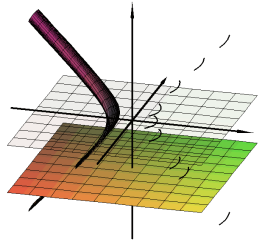}
\end{minipage}\hspace*{\fill}
\end{center}
\vspace{-1em}
\caption{The spacetime diagram of the magic field and its sources.}
\vspace{-1em}
\label{fig:Acc}
\end{figure}

The flat spacetime limit (i.e. $G\rightarrow 0$) of the charged rotating C-metric can be shown to lead to an electromagnetic field of two counter-rotating bent charged discs  undergoing uniform acceleration (see Fig. \ref{fig:Acc}, and \cite{BiKofAcc} for details).

\section{Removing the conical singularities}
The charged C-metric represents a pair of uniformly accelerated black holes with mass $m$,  charge $q$ and acceleration $A$. These can be ``immersed'' in an external electric field $E$ using the appropriate generating technique by Ernst. (The field breaks the asymptotic flatness.) For a suitably chosen value of $E$ the axis becomes regular everywhere. This is because by adding the external field we include the ``physical'' source of the acceleration in the solution -- the electromagnetic force. Utilizing the new factorized form of the C-metric, Ernst's simple ``equilibrium condition'' $mA = qE$ valid for small accelerations is generalized for an arbitrary $A$. See \cite{BiKofErnst} where rotation is also included.


\section*{References}
\bibliography{ae100prg-kofron}

\end{document}